\documentclass[11pt,a4paper,titlepage]{article}
\pdfoutput=1

\usepackage[utf8]{inputenc}
\usepackage{siunitx}
\usepackage{graphicx}
\usepackage{textcomp}
\usepackage{xcolor}
\usepackage{cite}
\usepackage[normalem]{ulem}

\begin{document}

\begin{center}
{\LARGE\bf Design of a 10 GHz minimum-B quadrupole permanent magnet
electron cyclotron resonance ion source}
\vspace{5mm}

{\Large T.~Kalvas$^\textrm{a,}$\footnote{Corresponding author, email: taneli.kalvas@jyu.fi}, O.~Tarvainen$^\textrm{b}$, V.~Toivanen$^\textrm{a}$ and H.~Koivisto$^\textrm{a}$}
\end{center}

\begin{enumerate}
    \item Department of Physics, University of Jyväskylä,\\
P.O.~Box 35 (YFL), 40014 Jyväskylä, Finland
\item UK Research and Innovation, STFC Rutherford Appleton Laboratory,\\
Chilton, OX11 0QX, United Kingdom
\end{enumerate}

\section*{Abstract}

This paper presents a simulation study of a permanent magnet electron cyclotron resonance ion source (ECRIS) with a minimum-B quadrupole magnetic field topology. The magnetic field is made to conform to conventional ECRIS with $B_\textrm{min}/B_\textrm{ECR}$ of 0.67 and a last closed magnetic isosurface of 1.86$B_\textrm{ECR}$ at 10 GHz. The distribution of magnetic field gradients parallel to the field, affecting the electron heating efficiency, cover a range from 0 to 13 T/m, being similar to conventional ECRIS. Therefore it is expected that the novel ion source produces warm electrons and high charge state ions in significant number. Single electron tracking simulations are used to estimate plasma flux distribution on the plasma chamber walls and to provide an estimate of the ion density profile at the extraction slit then used in ion optical simulations demonstrating high transmission through the low energy beam transport. The designed ion source is intended to study if the quadrupole field topology could produce high charge state beams in comparable intensities to conventional ECRIS and efficiently transport them through a low energy beamline, thus paving the way for a superconducting ARC-ECRIS using the same field topology. Furthermore, the prospects of the presented ion source design as an injector of a single-ended accelerator for ion beam analysis are discussed.

\section{Introduction}

Electron Cyclotron Resonance Ion Sources (ECRIS) are used for the production of intense beams of high charge state ions e.g.\ for nuclear and material physics research and applications. The origins of the ECRIS lie in the plasma fusion research of the 1960’s with open-ended mirror machines, the first ion source being conceived when Richard Geller installed a rudimentary beam extraction system to the \SI{3}{\giga\hertz} PLEIADE plasma generator \cite{Geller_book}. The first truly successful ECRIS, SUPERMAFIOS \cite{Geller_supermafios}, featuring two stages and, more importantly, a magnetic sextupole field superimposed on the solenoid field, was constructed in 1974. The superposition of the two fields forming so-called minimum-B structure satisfies three conditions that are understood to be important for high charge state ion production: (i) it houses a closed resonance surface where $\omega_{RF}=\omega_{ce}=\frac{eB}{m_e}$ (efficient electron heating), (ii) provides adequate mirror ratios for plasma confinement (long ion confinement time) and (iii) suppresses magnetohydrodynamic (MHD) instabilites due to the increase of the field strength in radial direction, i.e.\ $\frac{\partial B}{\partial r}>0$, corresponding to concave curvature of the field lines and subsequent MHD-stability.

The performance of ECR ion sources has improved dramatically over the past decades owing to improvements of the magnetic plasma confinement, increases in the microwave heating frequency and techniques to stabilize the plasma at high densities. At the same time the mechanical structure of the ion source has been simplified e.g.\ by abandoning the two-stage approach and adopting direct waveguide-based microwave coupling. Nevertheless, the basic concept, i.e.\ plasma heating and confinement in a minimum-B magnetic field formed as a superposition of solenoid and sextupole fields, has remained the same despite of successful experiments using solenoid field with quadrupole and octupole radial fields \cite{Tamagawa, Jongen}. 

The design of modern ECR ion sources is based on semi-empirical scaling laws, suggesting most importantly that the extracted current at the peak of the ion charge state distribution (CSD) scales with the microwave frequency squared \cite{Geller_frequencyscaling}, i.e.\
\begin{equation}
    I_\textrm{peak}\propto f_{RF}^2.
\end{equation}
\noindent
Improving the ion source performance by frequency scaling implies that the strength of the minimum-B magnetic field (solenoid and sextupole) is adjusted accordingly to fulfill the semi-empirical scaling laws \cite{Hitz_scalinglaws}
\begin{eqnarray}
    &B_\textrm{inj}/B_\textrm{ECR}=4\\
    &B_\textrm{rad}/B_\textrm{ECR}=2\\
    &B_\textrm{ext}\approx 0.9B_\textrm{rad}\\
    &B_\textrm{min}\approx 0.4B_\textrm{rad},
\end{eqnarray}
where $B_{\textrm{inj}}$, $B_{\textrm{rad}}$, $B_{\textrm{ext}}$ and $B_{\textrm{min}}$ are the fields at the injection, radial wall of the plasma chamber, extraction and B-minimum whereas $B_\textrm{ECR}\textrm{[T]}=f\textrm{[GHz]}/28$ is the resonance field for cold ($\gamma = 1$) electrons. The magnetic field scaling sets a practical limit of \SIrange[range-phrase = --, range-units = single]{18}{20}{\giga\hertz} for room-temperature (RT) ion sources based on electromagnetic coils and a permanent magnet sextupole. Thus, modern sources, such as VENUS, SuSI, RIKEN \SI{28}{\giga\hertz} ECRIS and SECRAL-II \cite{Leitner_VENUS, Machicoane_SUSI, Higurashi_RIKEN, Sun_SECRAL-II}, operating at frequencies higher than \SI{20}{\giga\hertz} rely on superconducting technologies. Building an ECRIS that fulfills the above scaling laws at \SI{56}{\giga\hertz} is feasible using the state-of-the-art Nb$_3$Sn superconducting wire \cite{Lyneis_56ghz}, whereas further increase of the frequency and magnetic field requires either accepting a substandard field strength or R\&D on innovative concepts, the latter being the topic of this paper.

In the 1960's magnetically confined fusion research was also carried out with devices utilizing another type of magnetic field topology. In these devices an ECR-heated plasma was confined in a minimum-B quadrupole field created by ``baseball''-shaped electromagnets (e.g.\ MFTF-B \cite{Porter_MTBF} and GAMMA10 \cite{Inutake_GAMMA10}). These devices were intended for reaching fusion conditions with light ions but suffered from unwanted by-products, namely high charge state impurity ions such as O$^{4+}$ \cite{Ikeda_oxygen}. Later in 1980's the Constance B experiment demonstrated the existence of hot electrons, kinetic instabilities and most importantly high charge state ions up to Ar$^{11+}$ in a large volume quadrupole mirror machine operating with several kW's of 10.5~GHz microwave power \cite{Garner, Petty}. In these experiments the high charge state ions were detected directly from the end loss plasma flux. In 2006 the JYFL ion source group demonstrated that such minimum-B quadrupole field structure based on two ``yin-yang'' electromagnets can be used as an ECR ion source connected to a low energy beamline \cite{Suominen_ARC-ECRIS}. The first low-cost prototype of the so-called ARC-ECRIS operating at \SI{6.4}{\giga\hertz} suffered from numerous compromises, e.g.\ the highest mirror ratio in the first prototype was only about 1.3 and the transport of the extracted beam through the low-energy beamline was severely restricted. Nevertheless, the prototype source produced high charge state ion beams with \SI{}{\micro\ampere}-currents at modest (\SI{30}{\watt}) power \cite{Suominen_ARC-ECRIS}. The highest detected charge state in these experiments was Ar$^{6+}$.

Since the first prototype was built and tested, the ARC-ECRIS concept has been advanced by a design study of the magnetic field structure of high-frequency RT and superconducting versions of the device \cite{Suominen_SC-ARC}. It was demonstrated that the original design can be simplified by replacing the two ``yin-yang'' coils with a single racetrack coil bent to a semi-circle, which alleviates the extraction of the ion beam. The design study showed that an RT magnetic field configuration allowing up to \SI{18}{\giga\hertz} operation with less than \SI{200}{\kilo\watt} power consumption is feasible using the ARC-concept \cite{Suominen_SC-ARC}. Even more importantly it was demonstrated that a superconducting version of the ARC-ECRIS operating at \SI{100}{\giga\hertz} with appropriate mirror ratios can be realized with existing superconducting wire (Nb$_3$Sn) \cite{Suominen_SC-ARC}. Thus, adopting the given magnetic field topology for high-frequency ion source (i.e.\ following the frequency scaling) could offer a path to push ECRIS performances beyond the state-of-the-art once the efficient extraction and transport of high charge state ion beams is properly demonstrated.

Construction of a high-frequency ARC-ECRIS carries an elevated risk. The plasma distribution and wall flux pattern of the ARC-ECRIS differ significantly from those of a conventional ECRIS. This has two important implications: (a) the semi-empirical magnetic field scaling laws guiding the design of conventional sources are not directly applicable and (b) the distribution of the plasma flux at the extraction favors a rectangular slit instead of a circular aperture to maximize the intensity of the extracted beam, which implies that the beam line should be equipped with quadrupole focusing to match the ion beam properties and ion beam optics \cite{Koivisto_ARC}. Therefore, further development of the ARC-ECRIS concept requires demonstrating that the unconventional quadrupole field topology (with adequate mirror ratios) equipped with a slit extraction is comparable to the superposition of solenoid and sextupole fields in terms of high charge state ion beam production. For this purpose we have designed a \SI{10}{\giga\hertz} permanent magnet (PM) CUBE-ECRIS, which has the same magnetic field topology with the ARC-ECRIS and therefore serves as a low-risk intermediate step in what could be a paradigm shift in ECRIS technology.  

This paper presents the physics design of the CUBE-ECRIS including the PM structure and resulting magnetic field, comparison of the magnetic field parameters (e.g.\ mirror ratios) to those of a typical conventional ECRIS and the ARC-ECRIS prototype, electron tracking simulations, beam extraction simulations and beam transport simulations. Microwave coupling simulations are omitted here as the exact cavity features including plasma damping effect are unknown and fine-frequency tuning will be applied experimentally to minimize waveguide losses as well as to optimize the high charge state production. The construction of the CUBE-ECRIS at JYFL (based on the simulations presented hereafter) has been fully funded and will commence in 2020. The prospects of the CUBE-ECRIS are discussed in Section~\ref{discussion}.

\section{Physics design of the CUBE-ECRIS}

\subsection{Magnetic field}

\begin{figure}[!htb]
\centering
\includegraphics[width=0.6\columnwidth]{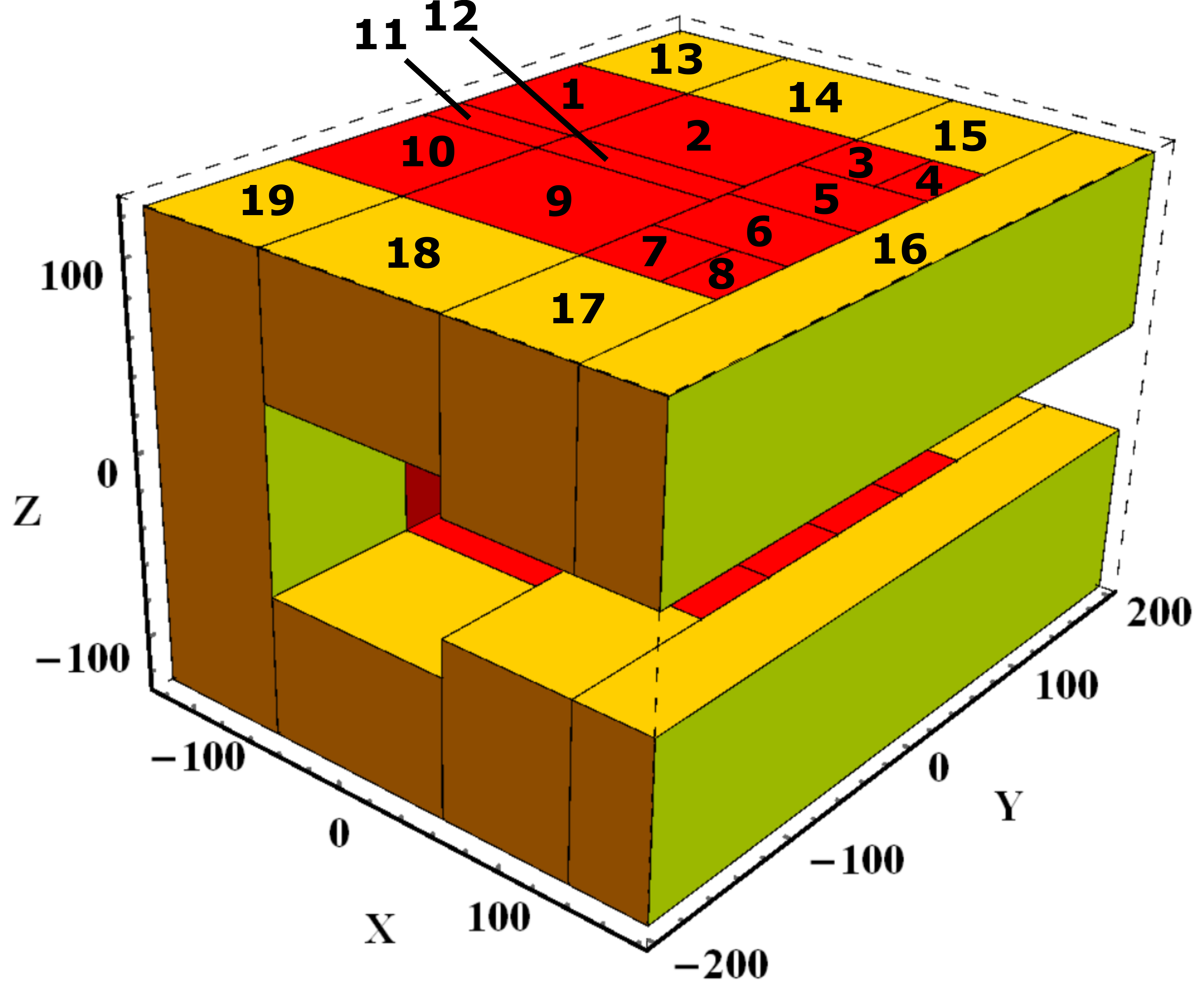}
\caption{The two-layer permanent magnet structure of the CUBE-ECRIS. The magnet numbering running clockwise through each layer and the coordinate system (in \SI{}{\milli\meter} units) defined in the figure are used throughout the manuscript. All magnets except \#1, \#10, \#11, \#13 and \#19 are mirrored about the $z=0$ plane.}
\label{CUBE-layers}
\end{figure}

The magnetic field design of the CUBE-ECRIS has been carried out with Radia3D software \cite{Radia3D}. The common thread of the design philosophy has been to use permanent magnet blocks with rectangular cross sections in all symmetry planes, i.e. asymmetric cubes, with the magnetization vector aligned along one of the edges (exceptions to this are outlined below). Such approach mitigates difficulties in assembling the PM array and reduces the cost by minimizing the amount of waste material. The magnetic material chosen for the prototype is NdFeB-45H with a remanence, $B_r$ of \SIrange[range-phrase = --, range-units = single]{1.32}{1.36}{\tesla} (nominal value of \SI{1.34}{\tesla} was used in the simulations) and an intrinsic coercive force, $H_{cj}$, of more than \SI{1353}{\kilo\ampere\per\meter}. The prototype design is based on two nested layers of permanent magnet blocks shown in Fig.~\ref{CUBE-layers} with the individual magnets listed in Table~\ref{magnet_table}. The inner layer alone is sufficient for \SI{6.4}{\giga\hertz} operation whereas the outer layer boosts the field strength to become viable for \SI{10}{\giga\hertz}. The PM array including the magnet sizes, locations and magnetization angles were optimized through a thorough iterative process. Only the final version chosen for commissioning is discussed hereafter.

\begin{table}[!htb]
\centering
\caption{CUBE-ECRIS magnet sizes, locations and magnetization vectors.}
\begin{footnotesize}
\begin{tabular}{cccc}
\hline
\hline
Magnet  & Size & Center  & Magnetization \\
\#       & ($x,y,z$) (mm) & ($x,y,z$) (mm) & direction \\
\hline
\multicolumn{4}{l}{Inner layer (\SI{6.4}{\giga\hertz})}\\
1 & 75 $\times$ 95 $\times$ 250 & (-107.5, 60.5, 0) & $-\hat{y}$ \\
2 & 108 $\times$ 95 $\times$ 75 & (-16, 60.5, 87.5) & $-\hat{y}$  \\
3 & 45 $\times$ 51.5 $\times$ 95 & (60.5, 82.25, 77,5) & $-\hat{y}$ \\
4 & 30 $\times$ 51.5 $\times$ 95 & (98, 82.25, 77.5) & $+\hat{z}$ \\
5 & 75 $\times$ 56.5 $\times$ 95 & (75.5, 28.25, 77.5) & $-\frac{\sqrt{3}}{2}\hat{x}-\frac{1}{2}\hat{y}$ \\
6 & 75 $\times$ 56.5 $\times$ 95 & (75.5, -28.25, 77.5) & $-\frac{\sqrt{3}}{2}\hat{x}+\frac{1}{2}\hat{y}$ \\
7 & 45 $\times$ 51.5 $\times$ 95 & (60.5, -82.25, 77,5) & $+\hat{y}$ \\
8 & 30 $\times$ 51.5 $\times$ 95 & (98, -82.25, 77.5) & $+\hat{z}$ \\
9 & 108 $\times$ 95 $\times$ 75 & (-16, -60.5, 87.5) & $+\hat{y}$ \\
10 & 75 $\times$ 95 $\times$ 250 & (-107.5, -60.5, 0) & $+\hat{y}$ \\
11 & 75 $\times$ 26 $\times$ 250 & (-107.5, 0, 0) & $+\hat{x}$ \\
12 & 108 $\times$ 26 $\times$ 75 & (-16, 0, 87.5) & $-\hat{z}$  \\
\hline
\multicolumn{4}{l}{Outer layer (\SI{10}{\giga\hertz})}\\
13 & 75 $\times$ 90 $\times$ 250 & (-107.5, 153, 0) & $-\hat{x}$ \\
14 & 108 $\times$ 90 $\times$ 75 & (-16, 153, 87.5) & $+\hat{z}$ \\
15 & 75 $\times$ 90 $\times$ 95 & (75.5, 153, 77.5) & $+\hat{z}$ \\
16 & 45 $\times$ 396 $\times$ 95 & (135.5, 0, 77.5) & $+\hat{z}$ \\
17 & 75 $\times$ 90 $\times$ 95 & (75.5, -153, 77.5) & $+\hat{z}$ \\
18 & 108 $\times$ 90 $\times$ 75 & (-16, -153, 87.5) & $+\hat{z}$ \\
19 & 75 $\times$ 90 $\times$ 250 & (-107.5, -153, 0) & $-\hat{x}$ \\
\hline
\hline
\end{tabular}
\end{footnotesize}
\label{magnet_table}
\end{table}

\begin{figure}[!htb]
\centering
\includegraphics[width=0.7\columnwidth]{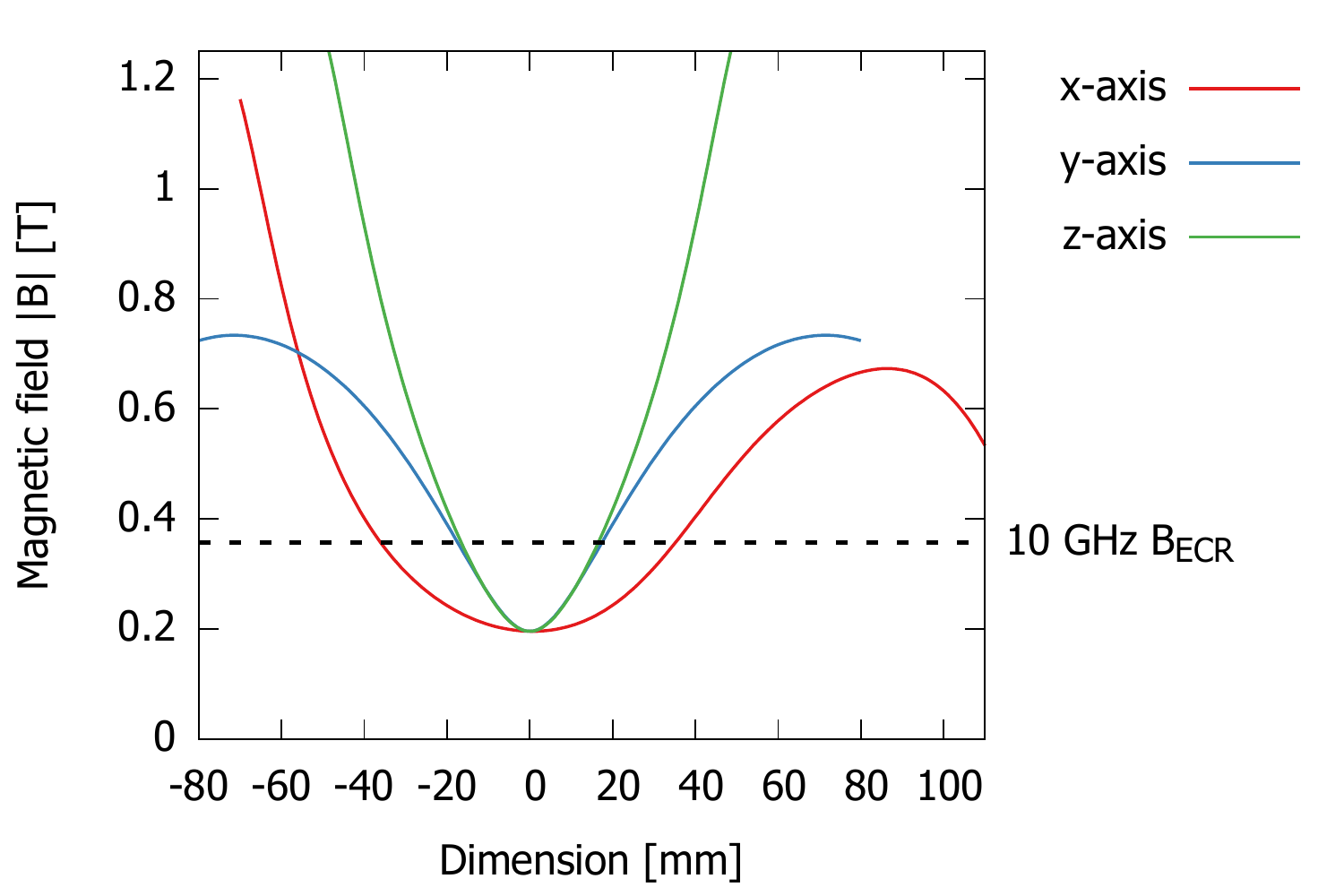}
\caption{The simulated total magnetic field, $B_{\textrm{tot}}$ of the CUBE-ECRIS along the lines parallel to each coordinate and axis passing through the minimum-B at ($x=0$~mm, $y=0$~mm, $z=0$~mm). The plot starts from the magnet surfaces in $-x$ and $\pm z$ directions.}
\label{CUBE_Bfield}
\end{figure}

The simulated total magnetic field, $B_{\textrm{tot}}$ of the CUBE-ECRIS along the lines parallel to each coordinate axis and passing through the minimum-B at ($x=0$~mm, $y=0$~mm, $z=0$~mm) are shown in Fig.~\ref{CUBE_Bfield}. The cold electron resonance field of \SI{0.357}{\tesla} for \SI{10}{\giga\hertz} microwave frequency is also marked in the figure for clarity. The minimum-B field of $0.67B_\textrm{ECR}$ for 10~GHz nominal frequency is chosen to be below the threshold of kinetic instabilities at $B_\textrm{min}=0.7-0.8B_\textrm{ECR}$, systematically observed with different ECR ion sources \cite{Tarvainen_ICIS15}.

The most important magnetic field parameters relevant for plasma (electron) confinement are summarized in Table~\ref{Bfield_parameters}. The mirror ratios\footnote{In the case of an ECRIS the mirror ratio is best defined as $B_{\textrm{max}}/B_{\textrm{ECR}}$ since the adiabatic invariance is violated each time the electron crosses the resonance zone.} in $-x$ and $\pm z$ directions depend on the thickness of the chamber wall and, thus, a range of values corresponding to \SIrange[range-phrase = --, range-units = single]{3}{5}{\milli\meter} wall thicknesses is given. In $\pm y$ direction the mirror ratio can be chosen freely, the optimum chamber wall location presumably being beyond the magnetic field maximum i.e.\ walls at $\left | y \right | >$ \SI{70}{\milli\meter}. Finally, in $+x$ direction the mirror ratio is defined by the location of the extraction slit as discussed below. For the values presented in Table~\ref{Bfield_parameters} it is assumed that the extraction slit is beyond the field maximum at $x=86$ mm. The $B_{\textrm{min}}$ value of \SI{0.24}{\tesla} corresponds to $B_{\textrm{min}}/B_{\textrm{ECR}}$ of 0.67 for \SI{10}{\giga\hertz}. It is worth noting that the $B_{\textrm{min}}/B_{\textrm{ECR}}$ and the mirror ratios can be affected by varying the microwave frequency.


\begin{table}[!htb]
\centering
\caption{CUBE-ECRIS magnetic field parameters for \SI{10}{\giga\hertz} operation. The range of values for $-x$ and $\pm z$ corresponds to \SIrange[range-phrase = --, range-units = single]{3}{5}{\milli\meter} chamber wall.}
\begin{tabular}{cccccc}
\hline
\hline
 Direction & $-x$ & $+x$ & $\pm y$ & $\pm z$ \\
\hline
$B_{\textrm{max}}$ & 1.07 -- 1.00 T  & 0.67 T & 0.73 T & 1.14 -- 1.07 T \\
$B_{\textrm{max}}/B_{\textrm{ECR}}$ & 2.97 -- 2.78 & 1.86 & 2.03 & 3.17 -- 2.97 \\
\hline
\hline
\end{tabular}
\label{Bfield_parameters}
\end{table}

\begin{figure}[!htb]
\centering
\includegraphics[width=0.5\columnwidth]{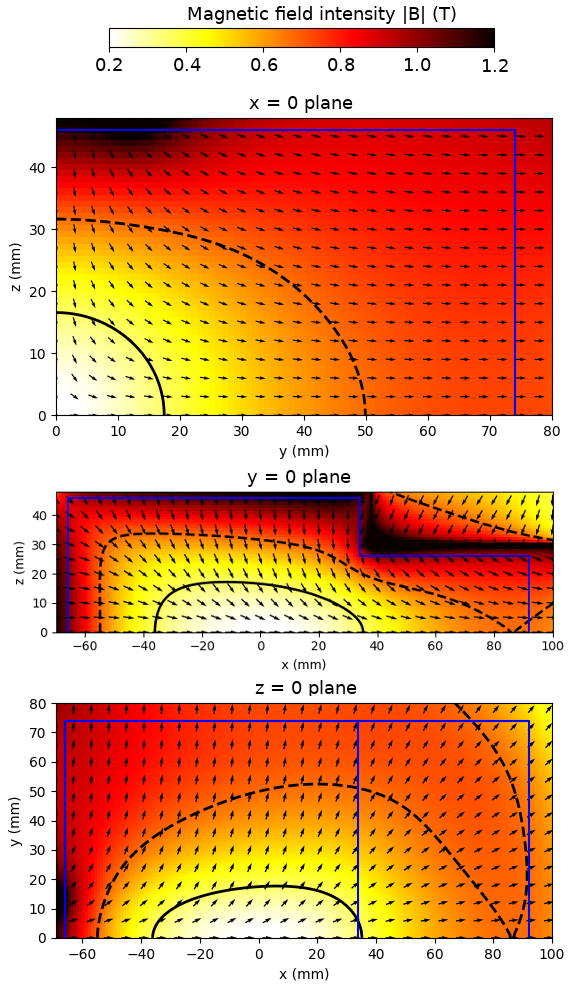}
\caption{The simulated magnetic field shown on three planes with $|B|$ presented with colormap and field direction with a quiver plot. The isosurfaces indicated by the black lines correspond to $B=B_\textrm{ECR}=0.357$ T (solid) and the last closed surface at $B=1.86B_\textrm{ECR}=0.673$ T (dashed). The blue lines represent the assumed plasma chamber surface.}
\label{fieldmap}
\end{figure}

Figure~\ref{fieldmap} shows the density and vector plots of the magnetic field in $x=0$~mm, $y=0$~mm and $z=0$~mm planes. The field is plotted in one quadrant/half of the structure. To obtain complete density and vector plots of the entire configuration the $x=0$ plot should be first mirrored about the $y=0$ plane and then the $z=0$ plane, the $y=0$ plot about the $z=0$ plane and finally the $z=0$ plot about the $y=0$ plane. The \SI{10}{\giga\hertz} ECR-surface $B=0.357$~T is indicated in each plot together with the last closed magnetic isosurface\footnote{The field strength on the last closed surface is defined by the location ($x$-coordinate) of the extraction slit.} where $B=0.673$~T. It has been argued that the field strength on the last closed surface should be approximately $2B_{\textrm{ECR}}$ \cite{Hitz_scalinglaws}, which is almost satisfied by the CUBE-ECRIS configuration for \SI{10}{\giga\hertz}. It is worth noting that most traditional ECR ion sources, where the magnetic field is a superposition of the solenoid and sextupole fields, fail to meet the condition of last closed magnetic isosurface being $2B_{\textrm{ECR}}$ or greater. This is because the radial component of the solenoid field counteracts the sextupole field at the chamber wall at three of the magnetic poles on the injection side and other three on the extraction side and creates the well-known radial plasma loss pattern (see e.g.\  \cite{Kalvas2014}).            

\begin{figure}[!htb]
\centering
\includegraphics[width=0.5\columnwidth]{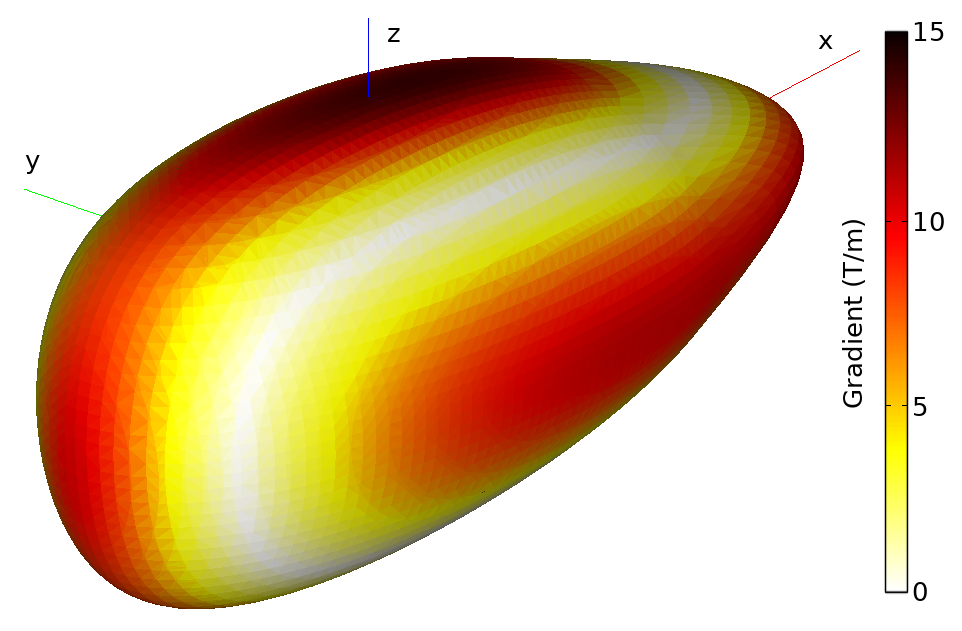}
\includegraphics[width=0.6\columnwidth]{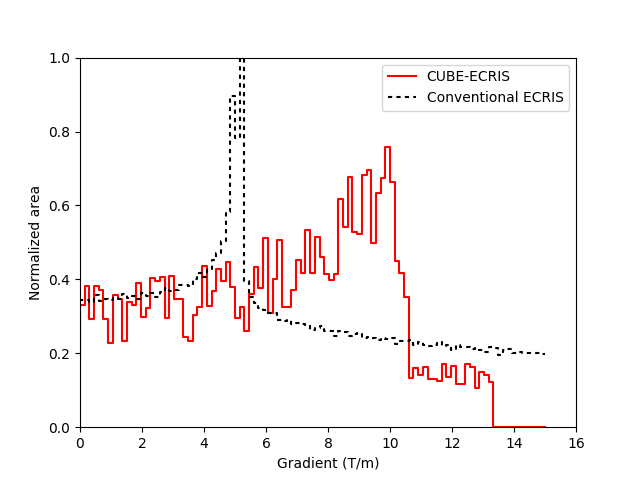}
\caption{The distribution of the magnetic field gradient parallel to the field, $|(\vec{B}/|\vec{B}|)\cdot \nabla \vec{B}|$, on the resonance surface presented as a three dimensional colormap and as a histogram representing the relative surface area corresponding to each gradient value. For comparison, the gradient distribution is also shown for a conventional ECRIS.}
\label{gradient_distribution}
\end{figure}

The efficiency of the electron heating, i.e.\ energy gain per resonance crossing, and the resulting electron energy distribution (EED) of the ECRIS plasma are believed to depend strongly on the local magnetic field gradient parallel to the field, i.e.\  $|(\vec{B}/|\vec{B}|)\cdot \nabla \vec{B}|$ \cite{Koivisto1999}. The EED affects the ionization rate, confinement properties and stability of the plasma against kinetic instabilities. Fig.~\ref{gradient_distribution} shows the distribution of $|(\vec{B}/|\vec{B}|)\cdot \nabla \vec{B}|$ on the \SI{10}{\giga\hertz} (cold electron) resonance surface of the CUBE-ECRIS enclosing a volume of 38 cm$^3$. A histogram plot comparing the gradient distributions of the CUBE-ECRIS and a conventional (solenoid + sextupole) minimum-B ion source, namely the JYFL \SI{14}{\giga\hertz} ECRIS \cite{Koivisto_ECR2}, is also shown. The gradient distributions of the two magnetic topologies are somewhat different but it can be expected that the electron heating properties in the CUBE-ECRIS are comparable to those of the conventional ion source. The two peaks in the histogram data of the conventional source at approximately 5~T/m are caused by the field gradient being almost constant near the chamber axis at the injection and extraction ends of the ECR surface. The effect is absent in the case of the CUBE-ECRIS.        

\subsection{Electron tracking simulations}

\begin{figure}[!htb]
\centering
\includegraphics[width=0.7\columnwidth]{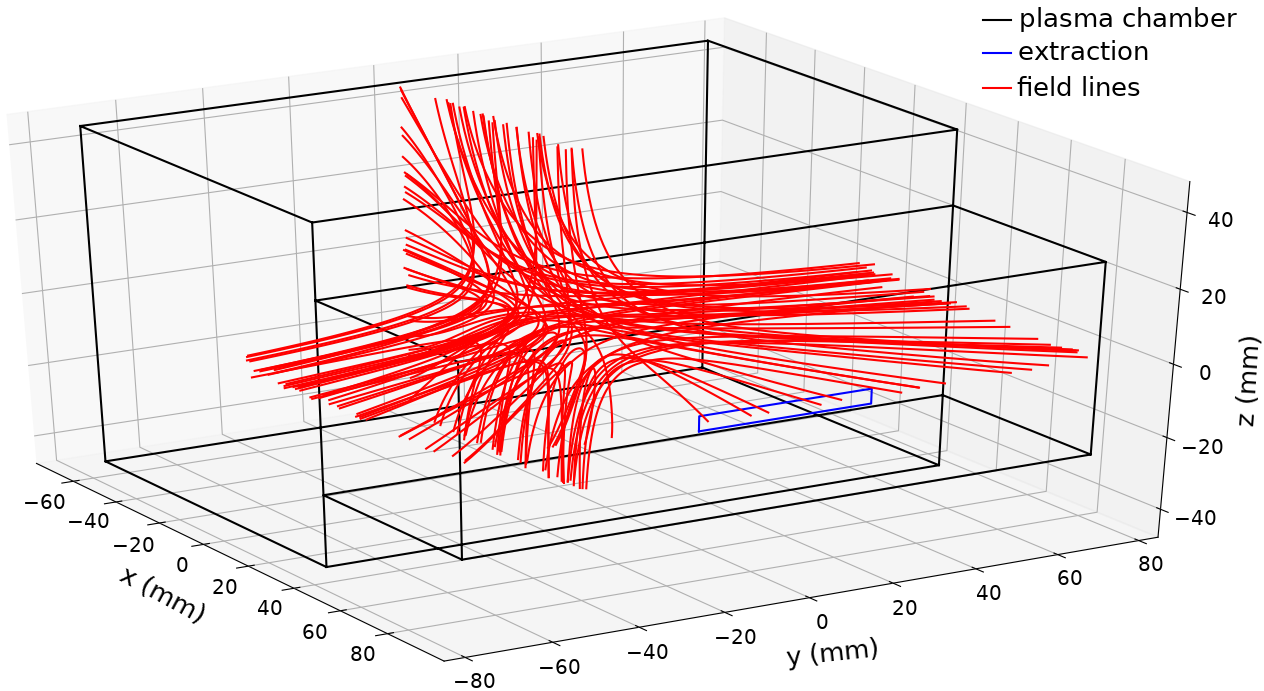}
\caption{A 3D view of the CUBE-ECRIS plasma chamber geometry and $N=100$ field lines passing through randomly selected points within the cold electron resonance surface and terminating on the chamber walls.}
\label{fieldlines_all}
\end{figure}

For designing the extraction of the CUBE-ECRIS and for evaluating the viability of the engineering choices, such as the chamber cooling, an estimate of the plasma flux distribution is needed. For this purpose it is sufficient to study the behaviour of electrons as it has been demonstrated with a conventional ECRIS that, despite being highly collisional and thus non-magnetized, the ions closely follow the electron flux \cite{Toivanen_collar,Carbon_cont}. An electron tracking code previously benchmarked (qualitatively) and then used for designing the HIISI \SI{18}{\giga\hertz} ECRIS \cite{Kalvas2014} is applied for this purpose. The code tracks single electron trajectories in the magnetic field of the ion source using a relativistic energy-conserving leapfrog-type algorithm. The RF fields and particle collisions are not considered. Since the electron energy distribution function is unknown, the electrons are launched isotropically from random locations within the $B_\text{ECR}$ resonance surface for relativistic electrons with kinetic energy $E_K$. If the particle is in the loss-cone of the magnetic bottle it will escape the confinement within a finite time. Otherwise it will remain confined in the plasma. The electrons are tracked until their trajectory intercepts with the plasma chamber or the maximum tracking time of 100 ns has elapsed in which more than 99 \% of the electrons within the loss-cone have escaped the confinement. Even though the method is a very rough approximation of the physics taking place in the ECR plasma the electron flux distributions on the plasma chamber walls provided by the simulation method have been shown to match well to experimentally observed plasma flux distributions as the magnetically confined electrons in an ECR mainly follow the field lines passing through the resonance surface. Collisional processes causing velocity space diffusion and cross-field spatial diffusion are expected to only slightly alter the electron distribution. Therefore, the field lines depicted in Fig.\ \ref{fieldlines_all} present a coarse view of the electron loss distribution.

The electron tracking simulations were completed with three electron energies: 10 keV, 100 keV and 200 keV. In all three cases the results are similar. On average about 70 \% of the electrons traced in the CUBE-ECRIS field are time-limited. These electrons are therefore outside the loss cone, i.e.\ confined. The fraction of confined electrons is similar to JYFL 14 GHz ECRIS as expected due to similar mirror ratios of the magnetic field. The escaping electron flux intercepts the chamber walls as illustrated in Fig.\ \ref{fieldlines_all}: On average, about 8 \% (of the escaping electrons) are lost to the wall at $x=-66$ mm, 54 \% to the walls at $|y|=74$ mm and 24 \% to $|z|=46$ mm. The fraction of electrons incident on the extraction electrode (positive $x$) depends slightly on the exact location, i.e.\ $x$-coordinate, of the electrode as shown in Table \ref{extloc}. It has been demonstrated with conventional ECRISs that optimizing the high charge state ion beam production requires the plasma electrode to be located at the maximum magnetic field or downstream from it \cite{Higurashi_plasmaelectrode1, Higurashi_plasmaelectrode2}. Thus, all extraction wall locations in Table \ref{extloc} are downstream from the magnetic field maximum at $x=86$ mm. The fraction of the escaping electron flux incident on a 4$\times$40 mm$^2$ extraction slit is approximately 4 \%, which is seemingly small but comparable to the 5 \% electron flux fraction incident on the extraction aperture of the JYFL 14 GHz ECRIS and exceeding the 2.5 \% fraction predicted for the 18 GHz HIISI ECRIS \cite{Koivisto_HIISI}, both these sources performing well.

\begin{figure}[!htb]
\centering
\includegraphics[width=0.7\columnwidth]{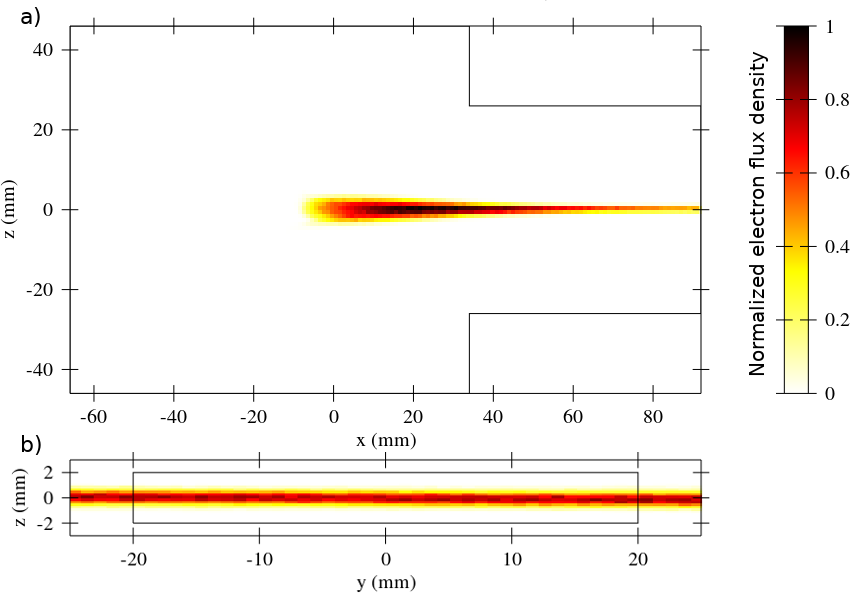}
\caption{The simulated $E_K=10$ keV electron flux distributions incident on a) the $y=-74$ mm wall and b) on the extraction wall ($x=92$ mm) with the $4\times 40$ mm$^2$ extraction slit marked by a rectangle.}
\label{flux}
\end{figure}

The electron flux densities incident on $y=-74$ mm chamber wall and on the proximity of the extraction slit located at $x=92$ mm (nominal) are presented in Fig.~\ref{flux}. The first plot serves to demonstrate that the waveguide delivering the 10 GHz power (and vacuum pumping on the opposite side) can be placed conveniently on the y-wall without overlapping with the escaping electron flux. The electron flux distribution on the walls or the extraction slit depends only weakly on the electron energy. The notable effect is that the width of the flux pattern increases with the electron energy similar to conventional ECRIS \cite{Kalvas2014}. The flux distribution within the extraction slit is approximately Gaussian with a standard deviation of 0.49 mm (in z-direction) for 10 keV electrons, 0.92 mm for 100 keV and 1.3 mm for 200 keV. Based on these simulations an extraction slit width of 4 mm has been initially chosen for the prototype design. With such a choice about 88 \% of the electron flux is located within the slit width in the 200 keV case. 

\begin{table}[!htb]
\centering
\caption{The simulated fraction of 10 keV electrons escaping towards the extraction wall (positive x) and through the 4$\times$40 mm$^2$ extraction slit.}
\begin{tabular}{cccc}
\hline
\hline
 Extraction plane & $x=87$ mm & $x=92$ mm & $x=97$ mm \\
\hline
Flux to $+x$ plane & 15.6 \% & 14.7 \% & 13.8 \% \\
Flux to slit &  4.43 \% & 4.06 \% & 3.74 \% \\
\hline
\hline
\end{tabular}
\label{extloc}
\end{table}

\subsection{Extraction and beam transport}

The ion optical design of CUBE-ECRIS extraction was carried out using IBSimu \cite{Kalvas_IBSimu}. In the simulations the flux of ions emerging from the plasma is assumed to follow the distribution of the electron flux incident on the plasma electrode nominally at the $x=92$ mm plane. The emitted flux is therefore uniform in the y-direction and has a Gaussian distribution in z-direction, with standard deviation of 0.49 mm corresponding to the simulated electrons with 10 keV energy. All emitted charge states are assumed to have the same spatial distribution as no experimental data contradicting the argument exist for the CUBE-ECRIS. It is acknowledged though, that higher charge states may be concentrated in the central plane of the ion source as observed for conventional ECRISs \cite{Wutte_emittance, Panitzsch_distribution}. The beam current of each charge state (in \textmu A) is taken from a Gaussian distribution with reasonable estimates for the mean charge state and standard deviation reflecting the performances of conventional sources (for argon, for example, a peak charge state of 7 and standard deviation of 2 were used). The total current of the extracted beam is expected to be in the order of 1 mA which has been used as a nominal value in the simulations. Otherwise, the plasma model parameters used in the simulations are similar to those used elsewhere for the ECRIS extraction design \cite{JINST_Toivanen}. The design of the following low energy beam transport (LEBT) line was made with PIOL \cite{PIOL} using third order matrices and the particle distributions from the extraction simulations.

\begin{figure}[!htb]
\centering
\includegraphics[width=0.7\columnwidth]{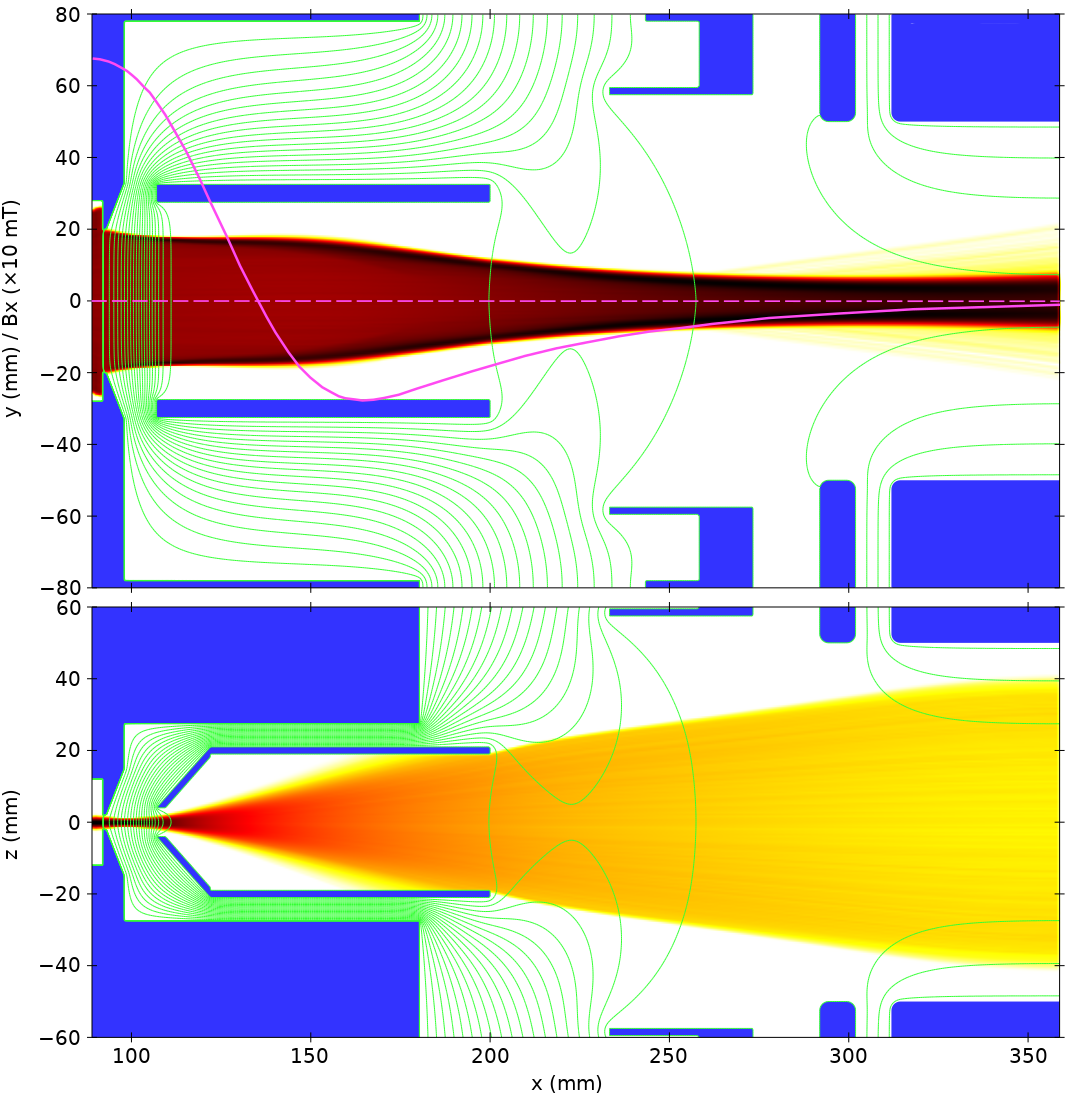}
\caption{Section view of the extraction geometry (electrodes in blue) with Ar$^{8+}$ ion trajectories projected to each plane and presented as a density of the trajectories. The magenta curve shows the x-component of the magnetic field on the y and z symmetry axis. The transmission of Ar$^{8+}$ through the puller electrode is 81 \%. After the extraction the beam enters the first electrostatic quadrupole of the transport line at $x=300$ mm.}
\label{geomext}
\end{figure}

The extraction design of the CUBE-ECRIS consists of slit-shaped plasma and puller electrodes accelerating the beam into the final energy in a single gap. Nominally the extraction is designed to operate with a 10 kV source potential, but based on simulations it is expected that the design works sufficiently well at least in the range from 7 kV to 14 kV. As the extraction takes place in the proximity of the magnetic field maximum, located well within the narrow gap of the PM structure at the ion source potential, the puller electrode must be long enough to shield the accelerated beam from the surrounding source potential. The puller electrode could have a nonzero voltage for additional adjustability, but in the presented design this feature is not used. If the adjustable puller is to be used the shape of the first grounded electrode (starting at $x=240$ mm) should be optimized for ion optics -- currently it is made as open as possible to allow for pumping. The lack of space near the extraction makes it challenging to achieve high beam transmission while maintaining sufficient clearances needed for the high voltage, and providing high conductance for pumping. The final design, presented in Fig.\ \ref{geomext}, is a functional compromise between several conflicting requirements.

\begin{figure}[!htb]
\centering
\includegraphics[width=0.5\columnwidth]{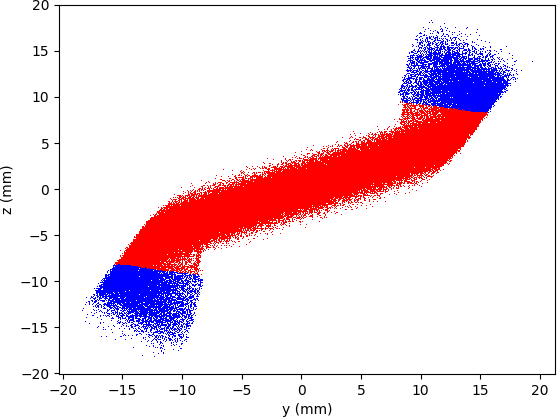}
\caption{$^{40}$Ar$^{8+}$ beam profile at x=150 mm in case of 10 kV source potential with trajectories collimated by the puller electrode drawn in blue and the rest in red.} 
\label{profile}
\end{figure}

\begin{figure}[!htb]
\centering
\includegraphics[width=0.5\columnwidth]{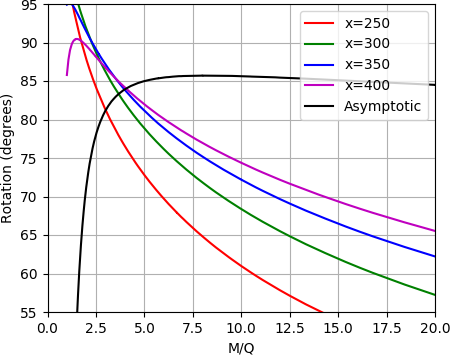}
\caption{Rotation of the ion beam due to the divergence of the magnetic field is a function of the mass-to-charge ratio for the nominal 10 kV ion source voltage.}
\label{rotation}
\end{figure}

The magnetic (fringe) field of the CUBE-ECRIS (presented in Fig.\ \ref{geomext}) is about 0.7 T at the extraction slit, pointing predominantly in the x-direction. As the extracted beam propagates through the region where the magnetic field diverges, the $B_z$ and $B_y$ components of the field cause rotation of the beam similar to a conventional ECRIS \cite{JINST_Toivanen} or a solenoid lens \cite{Kumar}. The rotation caused by the magnetic field is a linear effect in that it preserves the linear slit-shaped profile of the beam as long as the initial beam velocity is parallel to the x-axis. This is fulfilled for the most of the slit length as the focusing forces in the acceleration gap are weak compared to the magnetic force with the exception of the slit ends. Near the slit ends, at the proximity of $y=\pm 20$ mm, the focusing effect leads to an increased beam rotation, which can be seen as aberrations, for example, in the $^{40}$Ar$^{8+}$ beam profile at $x=150$ mm plane presented in Fig.\ \ref{profile}. The rotation caused by the magnetic field depends on the ion mass-to-charge ratio (or $M/Q$, where $M$ is the mass in atomic mass units and $Q$ is the charge state) and the ion source  voltage. The linear rotation in the extraction system was computed by tracing an ensemble of particles launched from a line in the center of the plasma electrode slit, ranging from  $(x,y,z)=(92,-15,0)$ mm to $(92,+15,0)$. The results are presented in Fig.\ \ref{rotation} for 10 kV ion source voltage and different distances from the ion source. The rotation induced by the magnetic field continues even after the magnetic field has diverged to zero. After a very long drift the rotation would reach the shown asymptotic value. The rotation is not very sensitive to the source voltage: In the case of $^{40}$Ar$^{8+}$, for example, at $x=300$ mm the beam rotation increases by $5.1^\circ$ compared to the nominal if the ion source voltage decreases to 7 kV (from 10~kV) and decreases by $5.0^\circ$ if the ion source voltage changes to 14 kV. The rotation has two important consequences for such extraction system. Firstly, the beam throughput depends on $M/Q$ due to collimation by the puller electrode --- for $^{40}$Ar, for example, the throughput drops almost linearly from 100 \% for charge state 1+ to 60 \% for 16+, being 81 \% for 8+ at 10~kV source potential. The throughput also varies as a function of the source potential. At 7 kV, for example, the throughput of $^{40}$Ar$^{8+}$ is 71 \% and at 14 kV, the throughput is 89 \%. The throughput can be increased by making the extraction slit narrower in the $y$-direction, but doing so would decrease the amount of extracted current and also increase the emittance of most of the beams of interest as the aberrations induced by the slit ends would not be collimated by the puller electrode as they are in the nominal case. The second consequence of the rotation is that the following beam transport line must be able to cope with the span of rotations of different particle species.

The fraction of ions collimated onto the puller electrode depends on the distance the beam must travel through the narrow electrode. Thus, the position of the plasma electrode affects the transmission efficiency assuming the plasma to puller electrode distance is constant. By multiplying the fraction of electrons escaping through the extraction slit (Table \ref{extloc}) with the transmission efficiency, a figure of merit for the expected beam current can be produced. For 10 kV source potential with the plasma electrode at $x=87$ mm the value is 3.22 \%, for $x=92$ mm it is 3.37 \% and for $x=97$ mm it is 3.42 \% for $^{40}$Ar$^{8+}$. Therefore it appears that it is beneficial to have the plasma electrode further downstream from the core plasma. On the other hand it is known that the position of the plasma electrode may affect the extracted charge state distribution \cite{Higurashi_plasmaelectrode1, Higurashi_plasmaelectrode2}, thus placing the plasma electrode far downstream from the field maximum might not be beneficial for the charge states of interest. Also it is known that the simulations used to calculate the electron fluxes do not model all the relevant physics. Therefore a compromise position $x=92$ mm, not far from the field maximum, has been chosen as the nominal position of the plasma electrode.

\begin{figure}[!htb]
\centering
\includegraphics[width=0.7\columnwidth]{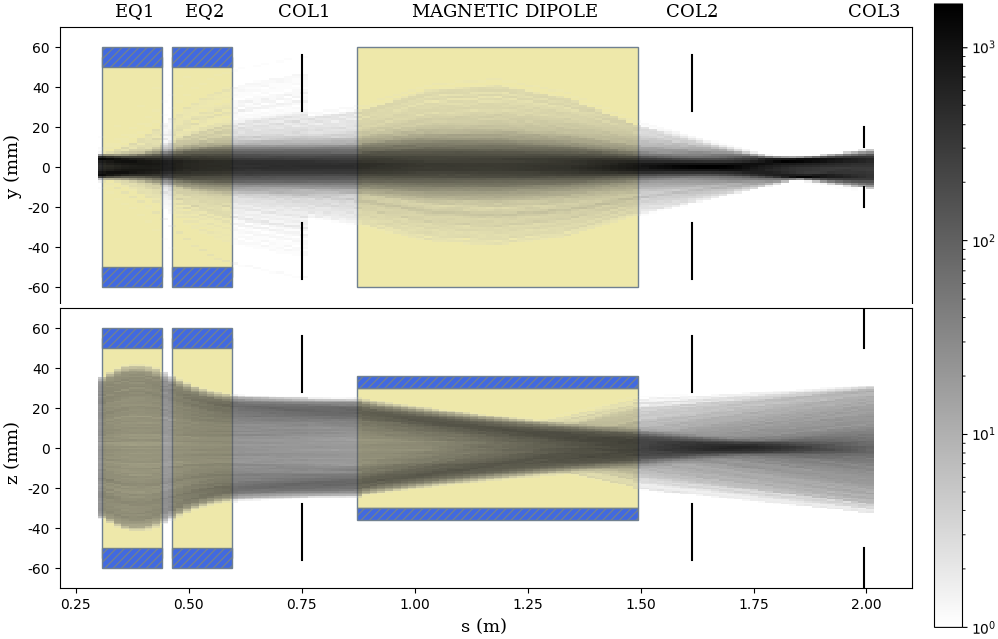}
\caption{$^{40}$Ar$^{8+}$ transport through the species separation into the beam intensity measurement after the rectangular COL3 collimator with 10 kV source potential. The coordinates $y$ and $z$ have the same orientation as in Fig.\ \ref{geomext}. The bending plane of the dipole is $z=0$.}
\label{lebt}
\end{figure}

\begin{figure}[!htb]
\centering
\includegraphics[width=0.5\columnwidth]{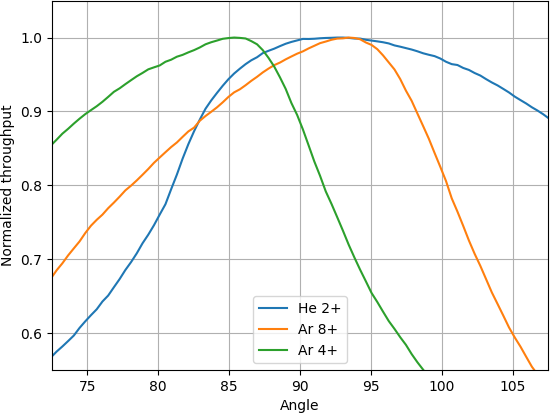}
\caption{Normalized throughput of the beam transport for $^4$He$^{2+}$, $^{40}$Ar$^{8+}$ and $^{40}$Ar$^{4+}$ beams as a function of the angle between the ion source and the transport line. The quadrupole voltages are kept constant at the optimized value for each of the beams.}
\label{rotopt}
\end{figure}

Transporting beams extracted from the CUBE-ECRIS requires quadrupole focusing to cope with slit shape of the beam. The solution chosen for the LEBT of the test bench planned for the CUBE-ECRIS consists of two 125 mm long electrostatic quadrupoles (EQ1 and EQ2) with 50 mm diameter bore and a pre-existing 102$^\circ$ magnetic dipole with a bending radius of 350 mm, pole gap of 70 mm and pole face angles of 33$^\circ$ for separation of beam species (see Fig.\ \ref{lebt}). The collimators COL1 and COL2 are round ISO-K63 flanges of the dipole chamber. The quadrupoles are oriented along the dipole bending direction to provide a tuning possibility for the two principal planes. The dipole bending should be done in the direction in which the beam is narrower and has a smaller projectional emittance to achieve appropriate separation of species. The beam is collimated by a rectangular $20\times 100$ mm$^2$ aperture (COL3) before measurement with a Faraday cup. As shown, the rotation of the beam depends on the distance from the ion source and, due to nonlinear effects, the beam profile is not a perfect rectangle. Therefore, it is not obvious what the orientation between the ion source and the LEBT should be. Thus, the system throughput was studied as a function of the rotation angle between the ion source plasma electrode / extraction slit (y-axis in Fig.\ \ref{geomext}) and the dipole bending plane. It was noticed that the optimum rotation angles for different ion beams are all close to 90$^\circ$ and that the throughput does not drop significantly in case of a slight deviation from the 90$^\circ$ (see Fig.\ \ref{rotopt}). Hence, it was concluded that the permanent magnet source could be mounted at a fixed 90$^\circ$ angle, which is very convenient as adjusting the rotation for different beams would be cumbersome. In a superconducting ARC-ECRIS the optimization of the high charge state ion beam intensities would involve varying the magnetic field strength and, therefore, an additional adjustment such as a tunable puller voltage would be required to compensate for the corresponding effect on the beam rotation.

The LEBT designed for the CUBE-ECRIS is capable of separating for example $^{129}$Xe charge states up to 30+ and 31+. The LEBT achieves transmission efficiency of $>90$ \% for $^{40}$Ar charge states $Q\ge 6$ and $>60$ \% for $Q\ge3$. About 10 \% of the ions entering EQ1 are lost for each charge state due to the aberrations near the slit ends. The total transmission efficiency from the ion source to the Faraday cup is presented in Table \ref{Ar_throughput} for selected ion species and source voltages together with the projectional normalized rms emittances of the beams at the start and the end of the LEBT. For comparison, the projectional normalized rms emittance of a $^{40}$Ar$^{8+}$ beam extracted from a conventional JYFL 14 GHz ECRIS is about 0.10 mm mrad in both planes \cite{JINST_Toivanen}.

\begin{table}[!htb]
\centering
\caption{The total transmission efficiency from the extraction aperture to the Faraday cup and normalized projectional rms emittances at $x=300$ mm and at the end of the LEBT in units of mm mrad for selected ion species and source voltages. At $x=300$ mm the projected emittances are calculated in the directions defined by the beam rotation (see Fig.\ \ref{rotation}) and labelled here as the emittance in \emph{wide} direction ($\epsilon_w$) and \emph{narrow} direction ($\epsilon_n$).}
\begin{tabular}{ccccccc}
\hline
\hline
Ion & Source & Through- & \multicolumn{2}{c}{At $x=300$ mm} & \multicolumn{2}{c}{At LEBT end} \\
    & (kV) & put (\%)       & $\epsilon_n$ & $\epsilon_w$ & $\epsilon_y$ & $\epsilon_z$ \\
\hline
$^{4}$He$^{2+}$   & 10 & 52  & 0.057 & 0.74 & 0.078 & 0.62 \\
$^{40}$Ar$^{11+}$ & 10 & 68  & 0.051 & 0.36 & 0.070 & 0.36 \\
$^{40}$Ar$^{8+}$  & 7  & 70  & 0.048  & 0.26 & 0.061 & 0.25 \\
$^{40}$Ar$^{8+}$  & 10 & 79  & 0.098 & 0.32 & 0.087 & 0.30 \\
$^{40}$Ar$^{8+}$  & 14 & 84  & 0.23  & 0.47 & 0.13  & 0.30 \\
$^{40}$Ar$^{4+}$  & 10 & 76  & 0.25  & 0.32 & 0.10  & 0.11 \\
\hline
\hline
\end{tabular}
\label{Ar_throughput}
\end{table}

In an application where only beams with the same or very close to each other M/Q values and a fixed ion source voltage were to be used, the LEBT design could be simplified. The extracted beam could be directly injected into a custom-made magnetic dipole with in-built quadrupole components for suitable focusing in both planes and species separation. Such system could also be perfectly rotated for the M/Q in question and, therefore, it would possibly achieve a slightly higher transmission efficiency than the system presented above.

\section{Discussion}
\label{discussion}

The simulation work described in the previous sections has demonstrated the feasibility of the CUBE-ECRIS concept (at least) up to \SI{10}{\giga\hertz} with adequate mirror ratios and appropriate beam transport through the low energy beamline. Hence, the engineering design of the prototype source and beamline is based on the presented concept. The primary goal of the prototype is to demonstrate the feasibility of the described magnetic field topology for adequate electron heating and high charge state ion (beam) production, therefore advancing the ARC-ECRIS concept. These will be assessed by measuring the plasma and wall (thick-target) bremsstrahlung spectra revealing the maximum electron energy and reflecting their energy distribution, detecting optical emission lines of Ar$^{10+}$ and Ar$^{13+}$ ions with a high resolution spectrometer \cite{Kronholm_possu1, Kronholm_possu2}, measuring the extracted beam currents from a Faraday cup and measuring the phase space distributions and thus emittances with an Allison scanner. The success of the CUBE-ECRIS and the prospects of the ARC-ECRIS will inevitably be defined through a comparison to existing PM ECR ion sources. The reported beam currents of different sources are listed in Table~S1 in the Supplementary Material \cite{Pantechnik_website, LTSun2008, Hitz_AIEP, Xie_BIE100, Scott, Rick, Muramatsu_Kei2, Muramatsu_Kei3, DECRIS-PM}. An especially interesting case is the BIE100 PM ECRIS \cite{Xie_BIE100} performing extremely well (reaching charge states such as Ar$^{17+}$) at \SI{12.75}{} and \SI{14}{\giga\hertz} for both of which it has lower mirror ratios ($B_{\textrm{max}}/B_{\textrm{ECR}}$) in all directions along the escaping flux and weaker extraction field than the CUBE-ECRIS. This suggests that the CUBE-ECRIS could potentially operate at higher frequencies than the \SI{10}{\giga\hertz} design value.

Besides potentially paving the way for the ARC-ECRIS, the CUBE-ECRIS has its own merit for specific applications. Most importantly, the relatively simple PM configuration (with access for miniature ovens and sputter samples) and the proposed electrostatic beam focusing scheme make it ideal for installation on a high voltage platform. The ability to produce several \SI{}{\micro\ampere} beams of high charge state ions at high voltage platform would  be very attractive for Ion Beam Analysis (IBA) where 1--10 particle nA currents with an energy spread of 10$^{-3}$ or better are sufficient. Such single-ended electrostatic accelerator could potentially provide an attractive alternative to tandem accelerators and their negative ion sources as described in Ref.\ \cite{Julin_ERDA}. This is highlighted in Table~\ref{IBA_table} listing the ion beams routinely used for two IBA-methods, namely Rutherford Backscattering Spectrometry (RBS) and time-of-flight Elastic Recoil Detection Analysis (TOF-ERDA), utilizing a tandem-type accelerator and the corresponding alternatives based on a high charge state ion source and a single-ended accelerator with 750~kV platform potential.

\begin{table}[!htb]
\centering
\caption{Ion beams used routinely in Ion Beam Analysis with tandem-type accelerators and their possible alternatives with a 750~kV single-ended electrostatic accelerator.}
\begin{tabular}{cccc}
\hline
\hline
IBA    & Beam       & Energy & Beam \\
method & (neg. ion) & [MeV] & (pos. ion) \\
\hline
RBS & $^4$He & 1.5\footnote{Minimum practical energy} & $^4$He$^{2+}$ \\
TOF-ERDA & $^{35/37}$Cl & 8 & $^{40}$Ar$^{11+}$  \\
TOF-ERDA & $^{79/81}$Br & 13 & $^{84}$Kr$^{18+}$  \\
TOF-ERDA & $^{127}$I & 18 & $^{129}$Xe$^{24+}$ \\
\hline
\hline
\end{tabular}
\label{IBA_table}
\end{table}

The positive ion beams listed in Table~\ref{IBA_table} can be produced with PM ECR ion sources as demonstrated in Table~S1 in the Supplementary Material whereas 750~kV platform has been used routinely e.g.\ at Fermilab \cite{Holmes_fermilab}. The listed positive ions could be produced simultaneously with a single ECRIS instead of using a plasma ion source coupled with an alkali-metal charge exchange cell (He$^-$) and a cesiated sputter ion source (Cl$^-$, Br$^-$ and I$^-$) for the production of the negative ion beams. The beam currents delivered by the ECRIS are typically several \SI{}{\micro\ampere}, allowing to reduce the energy spread of the ion beams by high-resolution magnetic spectrometer and slit system thus reaching the required resolution of the IBA-method.  

\section*{Acknowledgements}

This work has been supported by the Research Infrastructure Programme of the Academy of Finland and the Academy of Finland Project funding (N:o 315855).

\end{document}


\pagestyle{empty}

\begin{landscape}
\section*{Supplementary Material}

\begin{footnotesize}
\begin{longtable}[t]{lccc>{\columncolor{MyGray}}cc>{\columncolor{MyGray}}cc>{\columncolor{MyGray}}cc>{\columncolor{MyGray}}cc>{\columncolor{MyGray}}cc>{\columncolor{MyGray}}cc>{\columncolor{MyGray}}cc>{\columncolor{MyGray}}cc>{\columncolor{MyGray}}cc>{\columncolor{MyGray}}c}


\caption{\label{PM_source_comparison} Beam currents [\textmu A] of PM ECR ion sources.}  \\
\hline
\hline
Ion Source & Ref. & Freq. [GHz]  & Ion CS & H & He & C & N & O & Ne & Mg & Ar & Ca & Ti & Fe & Kr & Ag & Xe & Ta & Au & Pb & Bi & U  \\ 
\hline
NANOGAN & [38] & 10 & \\
 &  &  &  1+ & 1000 & 1000 & & & & & & 100 & & & & & & & & & & & \\
 &  &  &  2+ &   & 100 & & & & & & & & & & & & & &  & & & \\
 &  &  &  4+ &  & & & & & & & 140 & & & & & & & & 10 & & & \\
 &  &  &  6+ &  & & & & &  & & 45 & & & & & & & & 9 & & & \\
 &  &  &  8+ &  & & & & & & & 20 & & & & & & & 10 & 8 & & &\\
 &  &  &  9+  &  & & & & & & & 5 & & & & & & & & 6 & & &  \\
 &  &  &  12+ &  & & & & & & &  & & & & & & 10 & 10 & & & &\\
 &  &  &  14+ &  & & & & & & &  & & & & & & 5 & 5 & 2 & & & \\
\hline
NANOGAN 14.5 & [38] & 14.5  \\
 &  &  &  1+ & 1500 & 1500 & & & & & & 700 & & & & & & & & & & & \\
 &  &  &  2+ &   & 200 & & & & & & & & & & & & & & & & &\\
 &  &  &  4+ &  & & & & & & & 280 & & & & & & & & 20 & & &\\
 &  &  &  6+ &  & & & & & & & 100 & & & & & & & & 20 & & & \\
 &  &  &  8+ &  & & & & & & & 60 & & & & & & & 10 & 20 & & & \\
 &  &  &  9+  &  & & & & & & & 20 & & & & & & & & 15& & &  \\
 &  &  &  12+ &  & & & & & & & & & & & & & 20 & 20 & & & &\\
 &  &  &  14+ &  & & & & & & & & & & & & & 15 & 20 & 4 & & &  \\
\hline
SUPER- & [38, 39] & 14.5  \\
NANOGAN &  &  &  1+ & 2000 & 2000 & 500 & 1000 & 1000 & 1000 & &  1000 & & & & 1000 & 500 & & & & & & \\
 &  &  &  2+ &   & 1000 & 350 & 300 & 400 & 300 & & 350 & & & & & & & & & & & \\
 &  &  &  4+ &  & & 200 & 100 & 300 & 200 & & 250 & & & & & 250 & & & & & & \\
 &  &  &  6+ &  & & 3 & 10 & 200 & 160 & & 200 & & & & & 250 & & & & & & \\
 &  &  &  8+ &  & & & & & 25 & & 200 & & & & & 200 & 220 & & & & & \\
 &  &  &  9+  &  & & & & & & & 90 & & & & & 90 & & & & & &\\
 &  &  &  11+  &  & & & & & & & 30 & & & & 25 & 30 & & & & & & \\
 &  &  &  14+  &  & & & & & & & 1 & & & & 15 & & & & & & & \\
 &  &  &  20+ &  & & & & & & & & & & & 4 & 15 & 4 & 20 & 10 & & & \\
 &  &  &  23+ &  & & & & & & & & & & & & & 14 & 1 & & & & \\
 &  &  &  26+ &  & & & & & & & & & & & & & 5 & & 10 & 3 & & \\
 &  &  &  27+ &  & & & & & & & & & & & & & 1 & & 6 & 1 & &  \\
 &  &  &  30+ &  & & & & & & & & & & & & & & & 1 & & & \\
\hline
SOPHIE & [40, 39] & 12 -- 14.5 &  \\
 &  &  &  6+ &  & & & & 630 & & & & & & & & & & & & & & \\
 &  &  &  7+ &  & & & & 90 & & & & & & & & & & & & & & \\
 &  &  &  8+ & & & & & & & & 500 & & & & & & & & & & & \\
 &  &  &  9+ & & & & & & & & 310 & & & & & & & & & &  & \\
 &  &  &  11+ & & & & & & & & 90  & & & & & & & & & & & \\
 &  &  &  12+ & & & & & & & & 35 & & & & & & & & & & & \\
 &  &  &  20+ & & & & & & & & & & & & & & 52 & & & & & \\
 &  &  &  23+ & & & & & & & & & & & & & & 31 & & & & &  \\
 &  &  &  26+ & & & & & & & & & & & & & & 24 & & & & & \\
 &  &  &  27+ & & & & & & & & & & & & & & 18 & & & & & \\
 &  &  &  30+ & & & & & & & & & & & & & & 1 & & & & & \\
\hline
BIE100 & [41,42,43] & 12.75 + 14.5 \\
 &  &  &  4+ &  &  & 503 & 600 & & & & & & & & & & & & & & &  \\
 &  &  &  5+ &  &  & 507 & 525 & & & & & & & & & & & & & & & \\
 &  &  &  6+ &  &  & 120 & 500 & & & & & & & & & & & & & & & \\
 &  &  &  7+ &  &  &  & 81 & & & & & & & & & & & & & & &\\
 &  &  &  8+ &  &  &  &  & 440 & & & 260 & & & & & & & & & & & \\
 &  &  &  9+ &  &  &  &  & 265 & & & 263 & & & & & & & & & & &  \\
 &  &  &  10+ &  &  &  &  &  & & &  264 & & & & & & & & & & &  \\
 &  &  &  11+ &  &  &  &  & 64 & & & 240 & & & & & & & & & & &  \\
 &  &  &  12+ &  &  &  & & 19 & & & 213 & & & & & & & & & & & \\
 &  &  &  13+ &  &  &  &  & 3 & & & 142  & & & & & & & & & & & \\
 &  &  &  14+ &  &  &  &  &  & & & 80  & & & & & & & & & & & \\
 &  &  &  15+ &  &  &  &  &  & & & 38  & & & & & & & & & & & \\
 &  &  &  17+ &  &  &  &  &  & & & 5 & & & & & & & & & 7 & & \\
 &  &  &  18+ &  &  &  &  &  & & &  & & & & & & & & & 8 & & \\
 &  &  &  19+ &  &  &  &  &  & &  & & & & & & & & & & 9 & 50 &  \\
 &  &  &  20+ &  &  &  &  &  & &  & & & & & & & & & & 9 & 50 &  \\
 &  &  &  21+ &  &  &  &  &  & &  & & & & & & & & & & 10 & 46 &  \\
 &  &  &  22+ &  &  &  &  &  & &  & & & & & & & & & & 8 & 42 &  \\
 &  &  &  23+ &  &  &  &  &  & &  & & & & & & & & & & 8 & 35 &  \\
 &  &  &  24+ &  &  &  &  &  & &  & & & & & & & & & & 7 & 26 &  \\
\hline
LAPECR1 & [39] & 14.5 \\
 &  &  &  1+ &  & 5000 & & & & & & & & & & & & & & & & & \\
 &  &  &  2+ &  & 2500 & & 1700 & & & & & & & & & & & & & & & \\
 &  &  &  5+ &  &  & & 160 & & & & & & & & & & & & & & & \\
\hline
LAPECR2 & [39] & 14.5\\
 &  &  &  6+ &  &  &  & & 1000 & & & & & & & & & & & & & & \\
 &  &  &  7+ &  &  &  & & 130 & & & & & & & & & & & & & & \\
 &  &  &  8+ &  &  &  & & & & & 460 & & & & & & & & & & & \\
 &  &  &  9+ &  &  &  & & & & & 355 & & & & & & & & & & & \\
 &  &  &  11+ &  &  &  & & & & & 166 & & & & & & & & & & & \\
 &  &  &  12+ &  &  &  & & & & & 62 & & & & & & & & & & & \\
 &  &  &  14+ &  &  &  & & & & & 16 & & & & & & & & & & & \\
 &  &  &  16+ &  &  &  & & & & & 2 & & & & & & & & & & & \\
 &  &  &  17+ &  &  &  & & & & & 2 & & & & & & & & & & & \\
 &  &  &  19+ &  &  &  & & & & & & & & & & 84 & & & & & &  \\
 &  &  &  20+ &  &  &  & & & & & & & & & & & 85 & & & & &  \\
 &  &  &  23+ &  &  &  & & & & & & & & & & & 53 & & & & &  \\
 &  &  &  26+ &  &  &  & & & & & & & & & & & 40 & & & & &  \\
 &  &  &  27+ &  &  &  & & & & & & & & & & & 24 & & & & &  \\
 &  &  &  30+ &  &  &  & & & & & & & & & & & 5 & & & & &  \\
 &  &  &  31+ &  &  &  & & & & & & & & & & & 2 & & & & & 4 \\
\hline
LAPECR3 & [39] & 14.5\\
 &  &  &  1+ &  & 6000 & & & & & & & & & & & & & & & & & \\
 &  &  &  4+ &  & & 600 & & & & & & & & & & & & & & & & \\
 &  &  &  5+ &  & & 260 & & & & & & & & & & & & & & & & \\
 &  &  &  6+ &  & & & & 360 & & & & & & & & & & & & & & \\
 &  &  &  9+ &  &  &  &  &  & & & 120 & & & & & & & & & & & \\
 &  &  &  11+ &  &  &  &  &  & & & 50 & & & & & & & & & & & \\
\hline
KEI2 & [44] & 8 -- 10 \\
 &  &  &  1+ &  & 1000 & 335 & & 800  & & & & & & & & & & & & & & \\
 &  &  &  2+ &  & 540 & 550 &  & 815 & & & 445 & & & & & & & & & & & \\
 &  &  &  3+ &  &  &  &  & 580 & & & 295 & & & & & & & & & & & \\
 &  &  &  4+ &  &  & 530 &  & 385 & & & 225 & & & & & & & & & & & \\
 &  &  &  5+ &  &  & 60 &  & 210 & & & 135 & & & & & & & & & & & \\
 &  &  &  6+ &  &  &  &  & 77 & & & 80 & & & & & & & & & & & \\
 &  &  &  7+ &  &  &  &  &  & & & 75 & & & & & & & & & & & \\
 &  &  &  8+ &  &  &  &  &  & & & 88 & & & & & & & & & & & \\
 &  &  &  9+ &  &  &  &  &  & & & 30 & & & & & & & & & & & \\
 &  &  &  11+ &  &  &  &  &  & & & 2.5 & & & & & & & & & & & \\
\hline
KEI3 & [45] & 10 -- 18 \\
 &  &  &  1+ &  & 3100 & 145 & 587 & 405 & 188 & & & & & & & & & & & & & \\
 &  &  &  2+ &  & 1920 & 405 & 700 & 475 & 215 & & & & & & & & & & & & & \\
 &  &  &  3+ &  &  & 677 & 482 & 367 & 215 & & & & & & & & & & & & & \\
 &  &  &  4+ &  &  & 565 & 382 & 280 & 35 & & & & & & & & & & & & & \\
 &  &  &  5+ &  &  & 75 & 185 & 187 &  & & & & & & & & & & & & & \\
 &  &  &  6+ &  &  &  & 10 & 99 & 167 & & & & & & & & & & & & & \\
 &  &  &  7+ &  &  &  &  &  & 50 & & & & & & & & & & & & & \\
\hline
DECRIS-PM & [46] & 14 \\
 &  &  &  5+ &  &  &  &  &  & & 450 & & & & & & & & & & & & \\
 &  &  &  7+ &  &  &  &  &  & & 140 & & 110 & & & & & & & & & & \\
 &  &  &  8+ &  &  &  &  &  & & 40 & 926 & 175 & & 70  & & & & & & & & \\
 &  &  &  9+ &  &  &  &  &  & & 15 & 500 & 220 & 90 & 85  & & & & & & & & \\
 &  &  &  10+ &  &  &  &  &  & &  &  &  & 72 & 80  & & & & & & & & \\
 &  &  &  11+ &  &  &  &  &  & & & 210 & 158 & 60 & 55 & & & & & & & &  \\
 &  &  &  12+ &  &  &  &  &  & & & 154 & 58 & 23 &  & 160 & & & & & & &  \\
 &  &  &  15+ &  &  &  &  &  & & &  &  &  &  & 180 & & & & & & &  \\
 &  &  &  17+ &  &  &  &  &  & & &  &  &  &  & 125 & & & & & & &  \\
 &  &  &  19+ &  &  &  &  &  & & &  &  &  &  & 50 & & & & & & & \\
 &  &  &  20+ &  &  &  &  &  & & &  &  &  &  &  & & 77 & & & & & \\
 &  &  &  23+ &  &  &  &  &  & & &  &  &  &  &  & & 70 & & & & & \\
 &  &  &  26+ &  &  &  &  &  & & &  &  &  &  &  & & 50 & & & & &  \\
\hline
\hline
\end{longtable}

\end{footnotesize}
\end{landscape}